\documentclass[conference]{IEEEtran}
\IEEEoverridecommandlockouts
\usepackage{cite}
\usepackage{amsmath,amssymb,amsfonts}
\usepackage{float}                    
\usepackage{algorithm}                
\usepackage{algpseudocode}            
\usepackage{listings}
\usepackage{xcolor}

\usepackage{xcolor}
\usepackage{listings}

\usepackage{subcaption} 
\usepackage{url}
\lstdefinestyle{cppstyle}{
  language        = C++,
  numbers         = left,
  numberstyle     = \tiny\color{gray},
  stepnumber      = 1,
  numbersep       = 8pt,
  basicstyle      = \small\ttfamily,
  keywordstyle    = \color{blue}\bfseries,
  commentstyle    = \color{green!50!black},
  showstringspaces= false,
  breaklines      = true,
  frame           = single,
}

\usepackage{graphicx}
\usepackage{subcaption}
\usepackage{textcomp}
\usepackage{xcolor}
\def\BibTeX{{\rm B\kern-.05em{\sc i\kern-.025em b}\kern-.08em
    T\kern-.1667em\lower.7ex\hbox{E}\kern-.125emX}}
\begin{document}

\title{Implementation of Tensor Network Simulation TN-Sim under
NWQ-Sim\\
}

\author{\IEEEauthorblockN{Aaron C. Hoyt}
\IEEEauthorblockA{\textit{Physics} \\
\textit{University of Washington}\\
Seattle, USA \\
ahoyt@uw.edu}
\and
\IEEEauthorblockN{Jonathan S. Bersson}
\IEEEauthorblockA{\textit{Chemistry} \\
\textit{University of Washington}\\
Seattle, USA \\
jbersson@uw.edu}
\and
\IEEEauthorblockN{Sean Garner}
\IEEEauthorblockA{\textit{ECE} \\
\textit{University of Washington}\\
Seattle, USA \\
seangarn@uw.edu}
\and
\IEEEauthorblockN{Chenxu Liu}
\IEEEauthorblockA{
\textit{PNNL}\\
Richland, USA \\
chenxu.liu@pnnl.gov}
\and
\IEEEauthorblockN{Ang Li}
\IEEEauthorblockA{\textit{PNNL} \\
\textit{University of Washington}\\
Richland, USA \\
ang.li@pnnl.gov}

}

\maketitle

\begin{abstract}
Large-scale tensor network simulations are crucial for developing robust complexity-theoretic bounds on classical quantum simulation, enabling circuit cutting approaches, and optimizing circuit compilation, all of which aid efficient quantum computation on limited quantum resources. Modern exascale high-performance computing platforms offer significant potential for advancing tensor network quantum circuit simulation capabilities. We implement TN-Sim, a tensor network simulator backend within the NWQ-Sim software package that utilizes the Tensor Algebra for Many-body Methods (TAMM) framework to support both distributed HPC-scale computations and local simulations with ITensor. To optimize the scale up in computation across multiple nodes we implement a task based parallelization scheme to demonstrate parallelized gate contraction for wide quantum circuits with many gates per layer. Through the integration of the TAMM framework with Matrix Product State (MPS) tensor network approaches, we deliver a simulation environment that can scale from local systems to HPC clusters. We demonstrate an MPS tensor network simulator running on the state-of-the-art Perlmutter (NVIDIA) supercomputer and discuss the potential portability of this software to HPC clusters such as Frontier (AMD) and Aurora (Intel). We also discuss future improvements including support for different tensor network topologies and enhanced computational efficiency. TN-Sim code can be found on NWQ-Sim: \url{https://github.com/pnnl/NWQ-Sim/tree/tn_sim}
\end{abstract}

\begin{IEEEkeywords}
High-performance computing, Quantum Computing, GPU, tensor network
\end{IEEEkeywords}

\section{Introduction}

Quantum Computing is emerging as a powerful computing hardware specialized for a broad class of problems from combinatorial optimization to quantum simulation. The Noisy-Intermediate-Scale Quantum (NISQ) computing era has machines ranging from tens to thousands of physical qubits, where the depth of circuits able to be executed remains low due to error dominating from quantum gates and decoherence~\cite{b1}. Most known useful quantum computations require tens of thousands to millions of physical qubits with reliable error correction to be executed on a few hundred fault tolerant qubits. Tensor network simulations of quantum circuits are employed to mark the bounds on classical quantum computability, enable quantum circuit cutting algorithms to optimize quantum hardware usage, and optimize quantum circuit compilation~\cite{b18}. Tensor networks are frequently used to benchmark and reproduce quantum supremacy experiments, such as the work of Zlokapa et al.~\cite{b25}, which simulated Google's 2019 quantum supremacy experiment using classical tensor network methods.

There are many heterogeneous compute methods that utilize tensor networks to make quantum computing more efficient. Circuit cutting is an approach to breaking down quantum circuits to be more efficiently run on NISQ devices for a tradeoff of classical post processing~\cite{b13,b14}. Already circuit cutting solutions find better solutions through using heterogeneous computing methods rather than quantum hardware alone~\cite{b10,b11}. Circuit compilation can be optimized through the use of tensor networks. AQC-Tensor allows for the approximate compilation of quantum circuits using tensor networks, this can greatly reduce the number of layers in hardware while optimizing for a small cost in fidelity~\cite{b15}.

To date there are many tensor network quantum simulators available on HPC clusters. Some of the most prominent examples include the Tensor Network Quantum Virtual Machine (TNQVM) as well as cuTensorNet from NVIDIA~\cite{b3,b24}. Many of these simulators target particular GPU architectures or lack configurability for tensor contraction backends. 

In this work we aim to develop a configurable, adaptable and efficient tensor network quantum simulator that scales from local personal computer hardware up to various HPC cluster platforms regardless of GPU architecture. We achieve this by utilizing ITensor tensor contraction routines for local simulation and using Tensor Algebra for Many-body Methods (TAMM) library for distributed and parallelized quantum gate contraction\cite{b4, b5}. TAMM is the ideal choice for tensor contraction on HPC systems due to current built-in support for tensor contraction for NVIDIA, AMD, and Intel GPU architectures. TAMM also has extensible backend support for customizable data distribution and communication, and advanced indexing and data slicing capabilities in contrast to other available tensor algebra solvers.

We implement this within the NWQ-Sim framework for quantum simulation~\cite{b6,b7}. NWQ-Sim is specifically designed to simulate large scale quantum circuits. Currently, state-vector and density-matrix simulation are implemented, which are designed for large-scale simulations of high-depth circuits with a medium number of qubits (typically $<$ 50). Tensor network simulations, by contrast, allow for the execution of a large number of qubits (hundreds or more) on circuits where entanglement scales according to the area law. For high entanglement circuits tensor simulations can compress the bond dimension trading off fidelity for circuit execution speed.

It remains an important and unsolved task to develop high quality simulation software that is capable of running on many types of HPC clusters that are highly customizable backends for alternative memory management and parallel programming models. This enables the rapid prototyping of novel tensor network algorithms to push the classical quantum compute boundary. NWQ-Sim is also a part of the QIR Alliance which pushes for the simulation of of quantum circuits agnostic to quantum programming languages. 

\section{Background}
We review the use of tensor networks to simulate quantum circuits and introduce matrix product states (MPS) depicted in Figure 1. We describe the current state of matrix product states based quantum circuit simulators.
\begin{figure}
  \centering
  \includegraphics[width=0.9\linewidth]{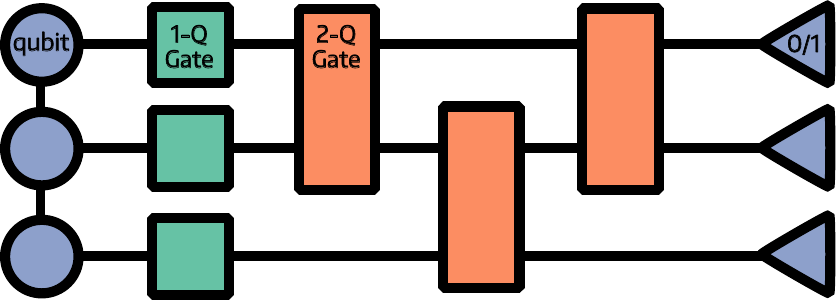}
  \caption{Depiction of MPS tensor train filled with rank 3 tensors, connected to a graph one and two qubit gates.}
  \label{fig:my-figure}
\end{figure}
\subsection{Quantum Circuit Simulation}
A quantum circuit can be directly represented as a tensor network. The initial state of the n-qubit system corresponds to a rank-n tensor.  A single qubit is a rank-1 tensor.  A 1-qubit gate and a 2-qubit gate are rank-2 and rank-4 tensors, respectively.  This correspondence allows for a quantum circuit to be simulated by directly contracting its n-qubits with all applied gates.
\begin{equation}
    Q^{mj} = \sum_{kli} U_i^{m(1)} U_{kl}^{ij(2)}Q^{kl}\label{eq1}
\end{equation}

Equation~\eqref{eq1} is a tensor contraction of a 2-qubit unitary quantum gate \(U^{(2)}\) with a quantum state, followed by a contraction with a 1-qubit unitary gate \(U^{(1)}\).  \(Q\) represents our tensor network. The result of the direct contraction method is a tensor where the entries are the amplitudes of the final state. However, the computational cost of this direct method scales exponentially with the number of qubits, making it impractical for larger systems. We, therefore, implement an MPS-based simulator.  This also provides the benefit of fine control over the level of approximation in our result.

An MPS, a.k.a, a tensor train, is a decomposition of a tensor based on the tensor singular value decomposition (SVD)~\cite{b23}.  
\begin{equation}
    Q = A^{q_1}_{i_1}A^{q_2}_{i_1, i_2}...A^{q_n}_{i_{n-1}}.\label{eq2}
\end{equation}
Each site \(A^{q}_{i,j}\) represents a qubit with one index that contracts with the network \(q\) and two (one if it is a terminal qubit) that represents the bond (tensor product) between neighboring qubits. 

This decomposition comes with the advantage that in circuits with low entanglement, the memory scaling is reduced.  The SVD (\(M = U \Sigma V^\dagger\)), in its connection with the Schmidt decomposition, tells us that non-zero singular values represent entanglement between qubits. The SVD is the best low rank approximation~\cite{b21}. Therefore, we can truncate singular values to control the size of the network, while simultaneously controlling the fidelity.  

Exactly expressing the quantum state with n qubits requires $d^n$ complex numbers in general, where d is the dimension of the local Hilbert space (d=2 for a qubit). However, if we know the entanglement is short-ranged, we can truncate the state to a fixed bond dimension $\chi$. Then the scaling becomes polynomial, $n * d * \chi^2$, rather than exponential.

The application of a single qubit gate becomes particularly simple in the MPS decomposition.  The gate is only applied to the tensor that represents its qubit and the rest of the network is not manipulated.
\begin{equation}
    A^{q'}_{i,j}=U^{q'(1)}_{q}A^{q}_{i,j}\label{eq3}
\end{equation}
The new MPS site replaces the old one without any effect to the other sites. The application of a local (the control and target qubit are adjacent) 2-qubit gate is also independent of the rest of the network, though it requires an SVD decomposition post contraction.
\section{Matrix Product State Quantum Circuit Simulation}
A circuit represented in the quantum assembly language (OpenQASM) is the input to the simulator. The gates are digested by NWQ-Sim's gate fusion process. This decomposes the circuit into arbitrary 1- and 2-qubit gates. The simulator controls the flow of gates to three categories of algorithms: 1-qubit gate application, local 2-qubit gate application, and non-local 2-qubit gate application. These functions apply gates to an MPS initialized in the zero state (\(|0...0\rangle\)). Finally, the network is sampled n-times, to emulate the shots of a physical quantum computer, with a measure all method. These algorithms are described here in more detail.
\subsection{Algorithms}
\begin{itemize}
    \item \textbf{One Qubit Gate}\\
    The $2\times2$ matrix representing a one-qubit gate is cast into the tensor type. A tensor contraction, in this case completely analogous to matrix multiplication, between the gate and the qubit is performed.
    \item \textbf{Two Qubit Gate}\\
    A variety of methods are known to perform local and non-local 2-qubit gate application in quantum circuit simulators \cite{b17,b19}.  The local method implemented in TN-Sim is a contract-decompose method using SVD, though the QR decomposition is substitutable. The control and target sites are contracted.  The gate and the sites are contracted. The resulting tensor is decomposed by SVD.  The left singular vectors are the control site.  The right singular vectors, with the singular values contracted in, are the target site.\\

    Non-local gates are more challenging in the MPS method, as the qubits between the control and target must also be manipulated. Common methods are the SWAP gate and the matrix product operator (MPO) based algorithms~\cite{b22}. Here we describe the linearly scaling (in qubit distance) bond propagation method implemented in TN-Sim ITensor backend for non-terminal qubits. \\

    We cast an arbitrary $4 \times 4$ matrix into a $2 \times 2 \times 2 \times 2$ tensor. We decompose this gate tensor by SVD, folding the singular values into the right singular vectors.  The left tensor is contracted with the control site, leaving us with a 4-index tensor, rather than the 3-index tensor of the MPS.  This extra index we identify as our propagating bond. It is the "link" between \(U\) and \(\Sigma\) in the SVD. This bond is to be contracted with the singular vectors of the gate tensor, but first we contract and decompose (by SVD) each intermediate qubit in succession. Once the final two qubits previous to the target qubit have been decomposed the propagating bond, \(\Sigma\), \(V^\dag\) and the target qubit are contracted and decomposed.  In this way the entanglement between the two qubits acted upon by the gate is correctly encoded.

    \item \textbf{Sampling}

    We implement the perfect sampling algorithm of \cite{b9}. In this method the marginal probability of each qubit is computed, the result is randomly chosen and the network is updated by projecting out the result and renormalizing.  This is repeated for the number of requested shots. The result is a vector of all possible output states, weighted by their probability. The sampling function is optimized by memoizing the networks corresponding to substrings of previous results.  These saved results are then used in place of recalculating the reduced density matrices at each step~\cite{b16}, effectively trading memory scaling for time scaling.
\end{itemize}

\section{Computational Methods}

To accommodate different computational scales, TN-Sim is implemented within NWQ-Sim using a dual-backend architecture. For local computation on single workstations, we use the ITensor library, which provides a comprehensive suite of built-in tools, including \verb|itensor::SpinHalf| for initialization, orthogonality functions like \verb|itensor::MPS::position|, and \verb|itensor::svd| for decomposition. For large-scale, distributed simulations on HPC platforms, we leverage the Tensor Algebra for Many-body Methods (TAMM) framework. TAMM is designed for performance at scale; while it includes a CPU-based SVD function from \verb|eigen|, we utilize NVIDIA's \verb|cusolverDnZgesvdj| to accelerate SVD with GPU hardware. Since orthogonality functions were not native to TAMM, they were implemented on top of the library for this work. Both backends, however, provide the core functionality of contracting arbitrarily ranked tensors according to Einstein notation.

\subsection{Implementations}

Quantum gates are parsed in NWQ-Sim originating from Quantum Assembly languages such as QASM, or directly called through the NWQ-Sim interface. To optimize the circuit execution during eventual simulation the initial quantum circuit is run through a gate fusion operation \verb|fuse_circuit_sv| provided in NWQ-Sim. This simplifies all gates down to a list of one and two qubit gates, furthermore any sequential gates with shared qubits will be compressed to decrease the total number of required contractions.

The tensor network is initialized with \(n\) sites, corresponding to \(n\) qubits. Each site tensor is a rank-3 tensor of dimensions \((1, d, 1)\), where \(d = 2\) is the local Hilbert space dimension of a qubit. The physical index at each site is initialized to represent the local \(\lvert 0 \rangle\) state, such that the full product state corresponds to \(\lvert 0 \rangle^{\otimes n}\). All bond dimensions are set to one, reflecting the absence of entanglement in the initial state.

\begin{figure}
  \centering
  \begin{minipage}{0.85\linewidth}
    \begin{lstlisting}[style=cppstyle]
// merge tensors at sites q0 and q1
IdxType Dl = bond_dims[q0];
IdxType Dr = bond_dims[q1 + 1];
tamm::Tensor<Cplx> M({
  bond_tis[q0],    // index l : left bond dimension at site q0
  phys_tis[q0],    // index p0: physical index of qubit q0
  phys_tis[q1],    // index p1: physical index of qubit q1
  bond_tis[q1 + 1] // index r : right bond dimension at site q1
});
M.set_dense();
M.allocate(&ec);

tamm::Scheduler sch{ec};
sch(
  M("l","p0","p1","r") =
    mps_tensors[q0]("l","p0","b") *
    mps_tensors[q1]("b","p1","r"),
  "merge two tensors",
  exec_hw
);
sch.execute(exec_hw);
    \end{lstlisting}
  \end{minipage}
  \caption{Merging two adjacent MPS site tensors into a single rank-4 tensor in TAMM.}
  \label{fig:mergetensors}
\end{figure}

The \verb|simulation_kernel| function loops over all one and two-qubit gates from \verb|fuse_circuit_sv|. We implement both local and non-local two-qubit gates. Local two-qubit gates in the TAMM backend are implemented by first merging the two-qubit MPS site, then contracting with the gate tensor using the \verb|scheduler|. Then the resulting tensor is decomposed using \verb|cusolverDnZgesvdj| where the bond dimension can be truncated. The SWAP non-local two-qubit gate is then implemented using swap gates to move each of the distant sites to become local, and is then reversed to the original positions after applying the two-qubit gate. An example of using the scheduler to merge two qubit sites is in Figure ~\ref{fig:mergetensors}.

Measurement operations \verb|MA_GATE| in TN-Sim are implemented through a perfect sampling algorithm~\cite{b9}. We first canonicalize the MPS in both left and right canonical forms to stabilize subsequent contractions using the \verb|left_canonicalize| and \verb|right_canonicalize| functions. Dynamic programming is utilized to speed up sampling by saving previously sampled bit string contraction outcomes.

\subsection{Parallelizing the TAMM Backend}

To take advantage of the inherent parallel nature of quantum circuit simulation we implement a task based parallelization strategy for the TAMM backend to execute multiple gates concurrently. As shown in Figure 3(a) we decompose the quantum circuit into layers of non-overlapping gates that can be executed in parallel. Many quantum circuits are heavily parallelizable such as Hamiltonian trotterization with local interactions and can be greatly accelerated with parallelized gate contraction. While many parallelization strategies rely on index slicing, we decide to use a DAG with distributed work assignment. We implement a hybrid architecture where TAMM manages shared global memory via global arrays and each rank pulls data to perform one and two qubit gates pulled from the DAG. We show that this speeds up quantum circuits with dense circuit layers but becomes dominated by synchronization time as bond dimension becomes too large. The advantage of this hybrid architecture is when tensor contractions no longer fit within GPU memory it is easy to begin index slicing to allocate multiple gpu ranks to gate contraction.

The layerization algorithm is implemented by iterating through the gate list provided from \verb|fuse_circuit_sv| and appending gates to layers with mutually exclusive sets of qubits. For simplicity, non-local two qubit gates are decomposed into local swap gates with a final local two-qubit gate. This layering pre-computation step ensures that all gates can be computed independently and in parallel.

Once the circuit is layered, the execution of gates is distributed among the available ranks using the dynamic work-stealing model shown in Figure 3(b) \cite{b29}. The \verb|AtomicCounterGA| within \verb|TAMM| manages the work to be distributed, each rank fetches an index to claim a gate for computation. This method ensures a balanced workload so that all ranks will be utilized during parallelization.

Each layer requires synchronization with the shared memory, implemented in TAMM with \verb|GlobalArrays|. After a layer is finished, the new tensor data with the updated bond dimension, are gathered from all ranks to collectively update the global MPS to ensure the state is consistent for the next layer of computation. This model allows TN-Sim to utilize distributed HPC resources with scaling qubit number, ranks, and bond dimension until synchronization time dominates.

\begin{figure}
  \centering

  \begin{subfigure}[t]{0.8\linewidth}
    \centering
    \includegraphics[width=\linewidth]{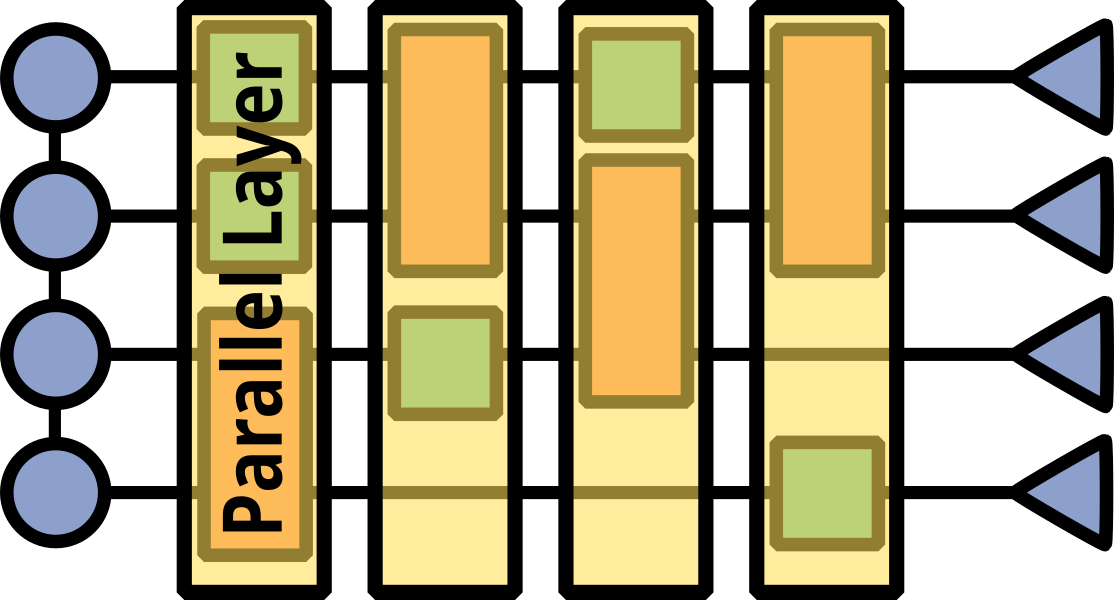}
    \subcaption{}%
    \label{fig:Parallel}
  \end{subfigure}

  \vspace{1em}

  \begin{subfigure}[t]{0.8\linewidth}
    \centering
    \includegraphics[width=\linewidth]{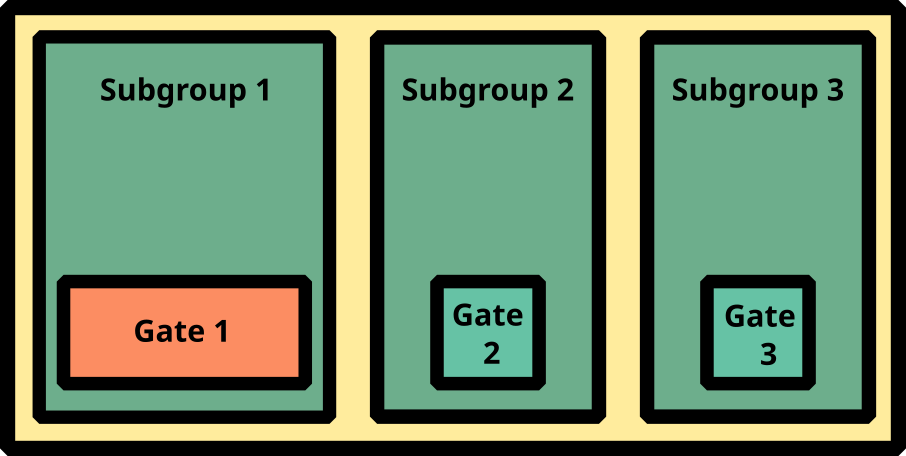}
    \subcaption{}%
    \label{fig:Subprocesses}
  \end{subfigure}

  \caption{(a) Layer decomposition of tensor-network quantum gates into non-overlapping qubit layers. (b) Subgroups executing gates in parallel in a non-overlapping qubit layer.}
  \label{fig:CombinedParallelSubprocesses}
\end{figure} 

\subsection{Portability}

NWQ-Sim is originally designed to run on a wide variety of computing architectures, with further support from local compute all the way to supercomputer HPC platforms. We follow this design philosophy when developing TN-Sim for NWQ-Sim. The TAMM library employs a unified GPU architecture design to support a variety of platforms. The experiments demonstrated in this paper utilized NVIDIA A100 nodes on Perlmutter.

\section{Evaluation}

The TN-Sim backend is validated against the NWQ-Sim test circuit suite QASM Bench for correctness \cite{b28}. Since distributed tensor contraction with TAMM comes with MPI overhead, we probe the regimes of utilizing the local backend of TN-Sim vs the distributed TAMM backend. We explore promising results of parallelizing TAMM gate contraction operations to separate subgroups. We also demonstrate the efficiency of the MPS simulation with a GHZ state.

\subsection{GHZ Demonstration}

Tensor networks excel at simulating quantum circuits which follow area-law entanglement~\cite{b12}. A fundamental example of such a circuit is GHZ state preparation, which will only require a maximum bond dimension of $\chi = 2$ throughout the entire chain in MPS representation. As the number of qubits scales, the memory requirements to represent the state only scale linearly as $O(n\chi^2)=O(4n)$. On the other hand, statevector simulation scales in time and memory with the full Hilbert space on the order $O(2^n)$. 

\begin{equation}
\lvert \mathrm{GHZ} \rangle = \frac{1}{\sqrt{2}} \left( \lvert 0 \rangle^{\otimes n} + \lvert 1 \rangle^{\otimes n} \right)
\end{equation}

\begin{figure}
  \centering
  \includegraphics[width=0.9\linewidth]{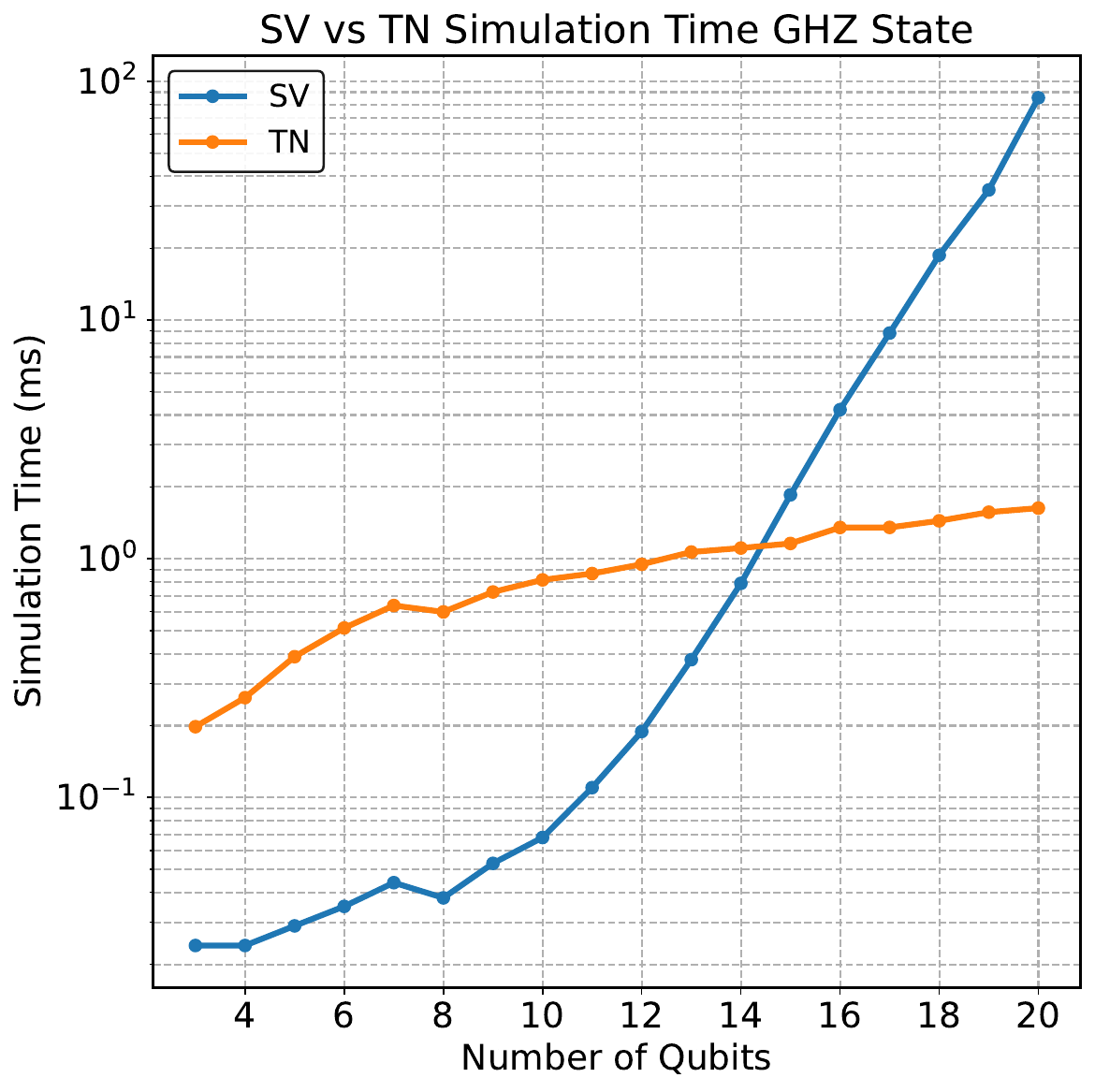}
  \caption{ITensor TN-Sim backend versus the statevector method used by SV-Sim. Simulation time over circuit width on the tensor network simulator scales near-linearly, while state-vector simulation times grow exponentially.}
  \label{fig:GHZ}
\end{figure}

\subsection{Large Width Circuits}

\begin{figure}[b]
  \centering
  \includegraphics[width=0.9\linewidth]{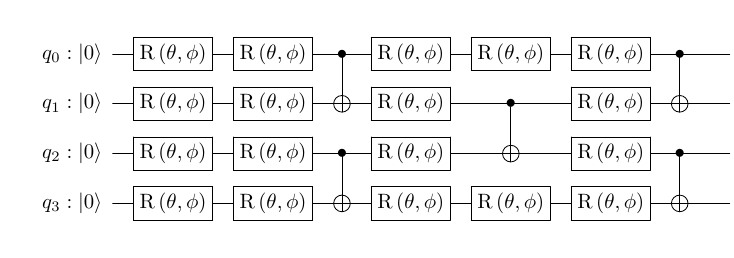}
  \caption{Circuit diagram for a brickwork state. Alternating between single and two qubit gates this circuit structure is nearly maximally dense per layer, which can be readily parallelized in our task based parallelism.}
  \label{fig:MergeParallel}
\end{figure}

\begin{figure}
  \centering
  \includegraphics[width=0.9\linewidth]{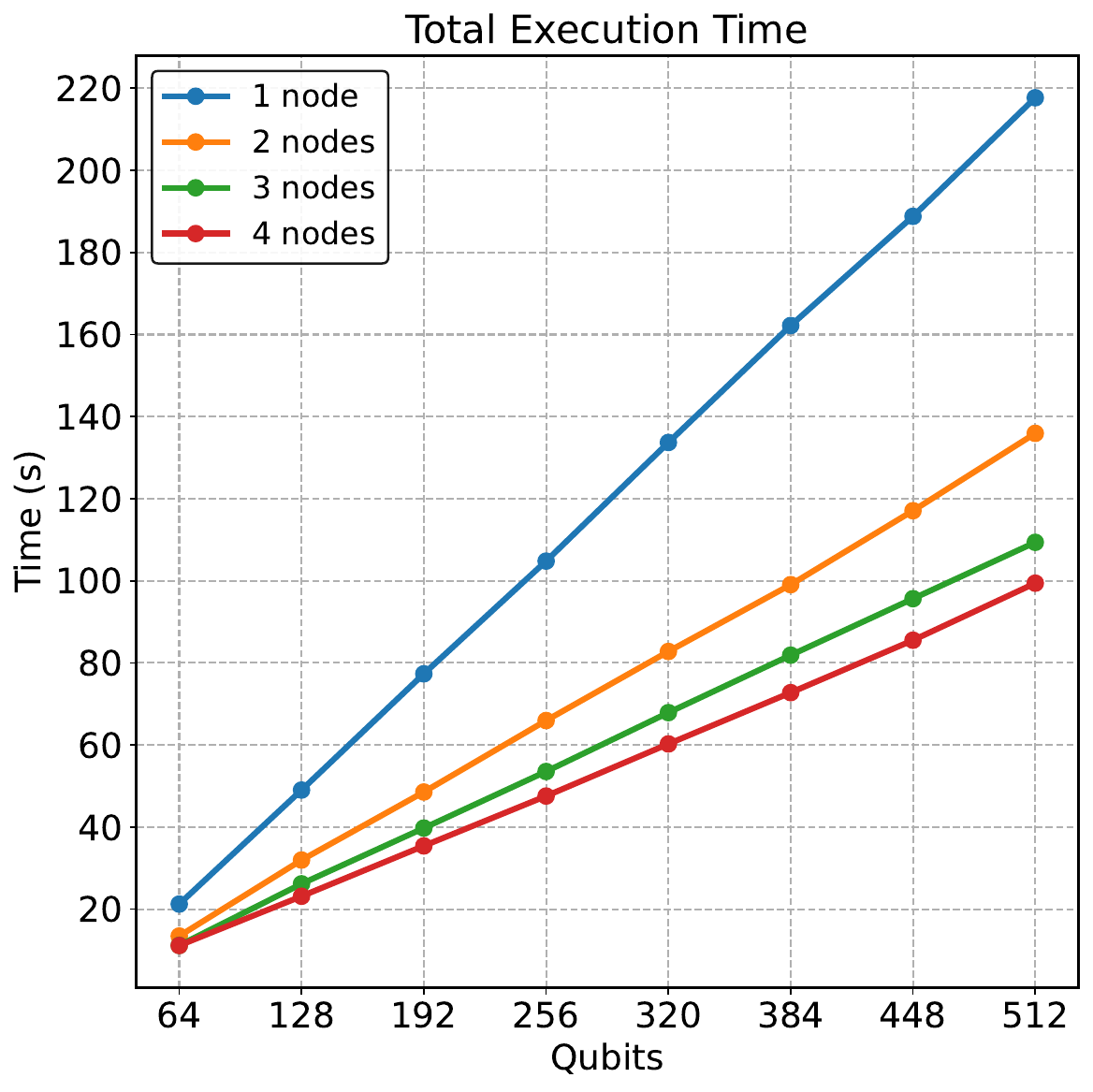}
  \caption{Total simulation time of executing a brickwork state with a fixed bond dimension of 250 and 80 layers with the TAMM backend. The total time scales linearly in qubit count as expected.}
  \label{fig:MergeParallel}
\end{figure}

To evaluate the performance and scalability of the task parallelized TAMM backend, we utilize a brickwork state benchmark. The structure of this circuit, depicted in Figure 5, is characterized by a nearly maximal density of non overlapping gates in each layer. This makes it an ideal test case for our parallelization strategy. The performance on this benchmark indicates the expected efficiency for other highly parallelizable quantum algorithms, such as trotterized Hamiltonian simulations with local interactions, Quantum Approximate Optimization Algorithms with local cost Hamiltonians, and Variational Quantum Eigensolvers employing local ansatzes.

\begin{figure*}[!htbp]
    \centering
    \begin{subfigure}[t]{0.32\linewidth}
        \centering
        \includegraphics[width=\linewidth]{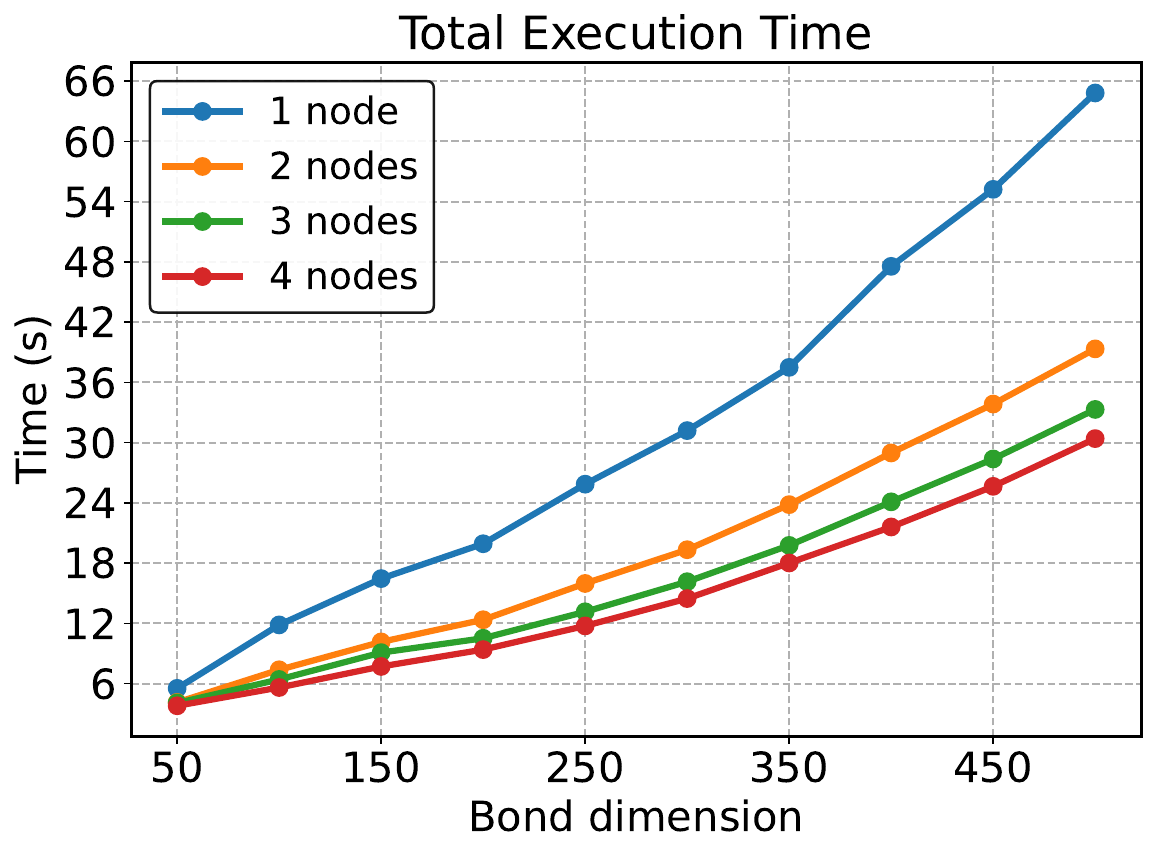}
        \caption{}
        \label{fig:MergeParallel1}
    \end{subfigure}
    \hfill
    \begin{subfigure}[t]{0.32\linewidth}
        \centering
        \includegraphics[width=\linewidth]{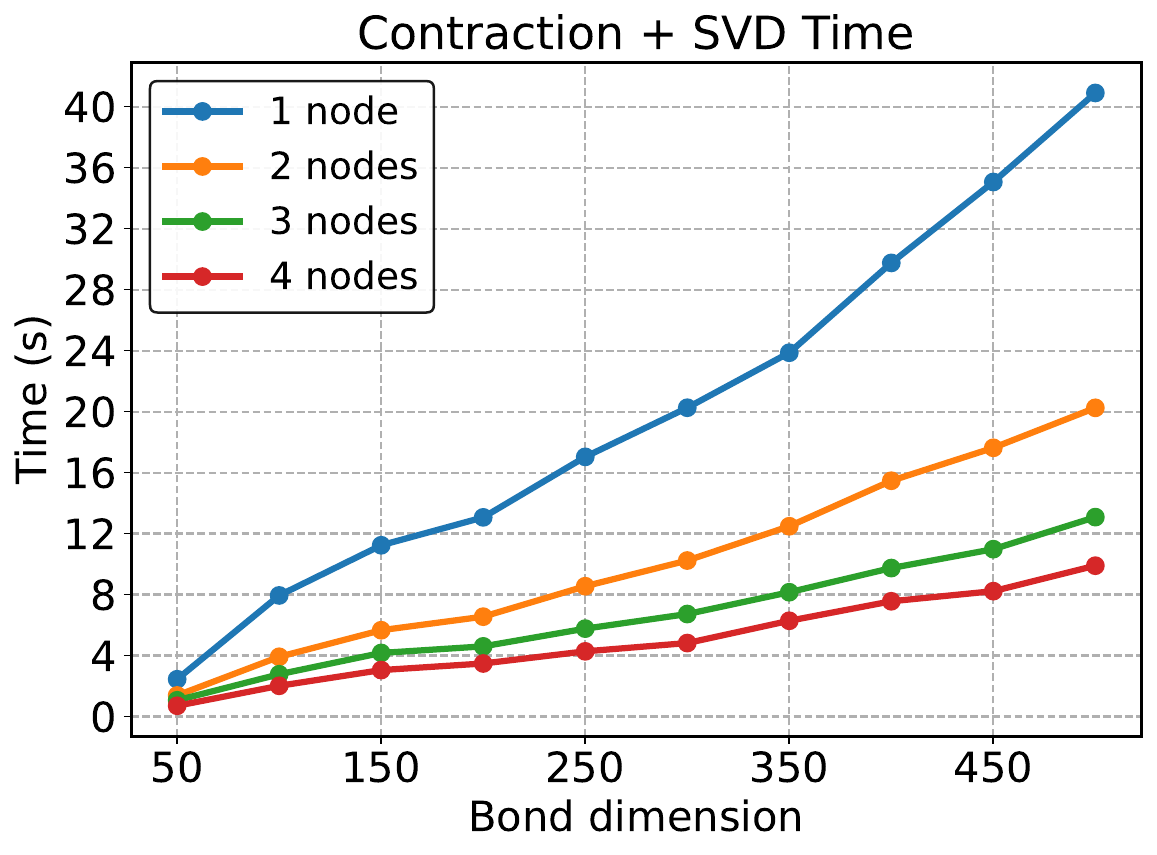}
        \caption{}
        \label{fig:MergeParallel2}
    \end{subfigure}
    \hfill
    \begin{subfigure}[t]{0.32\linewidth}
        \centering
        \includegraphics[width=\linewidth]{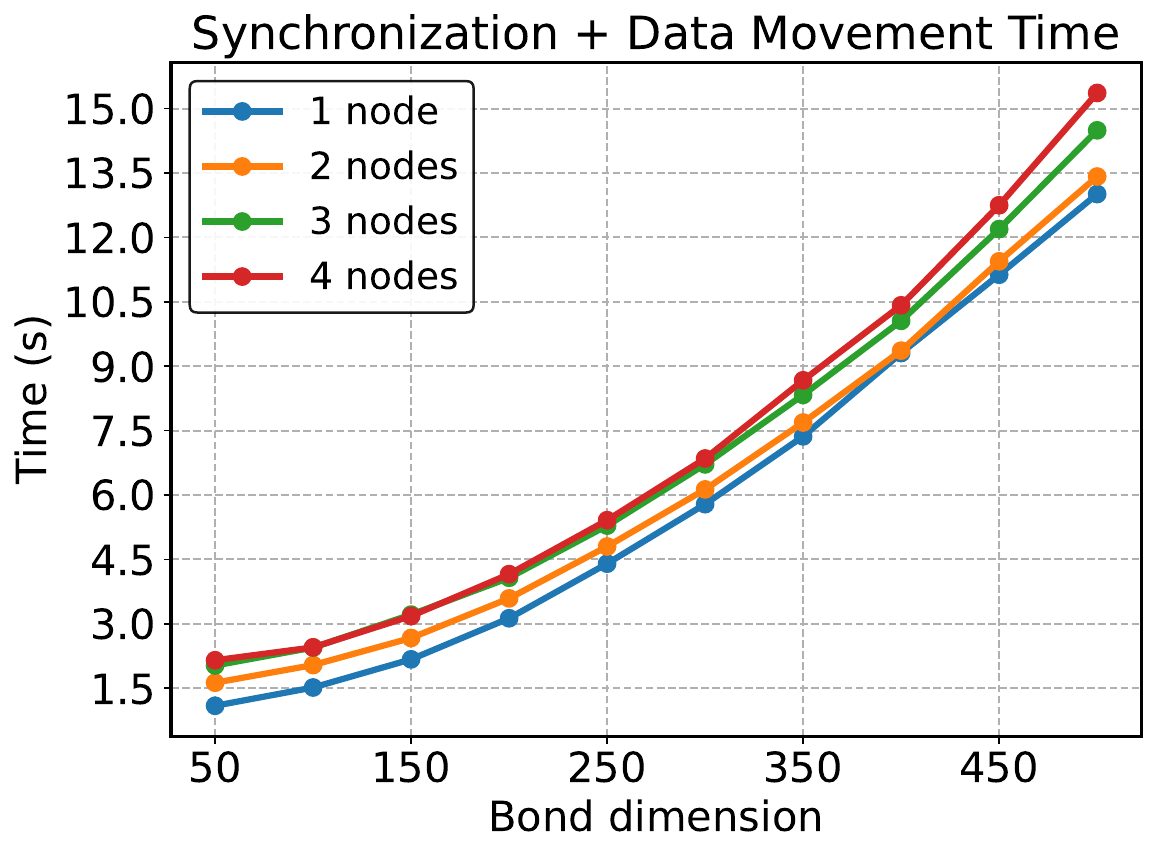}
        \caption{}
        \label{fig:MergeParallel3}
    \end{subfigure}

    \caption{
        Simulation time scaling for a 64 qubit brickwork circuit with depth $80$ as a function of the maximum bond dimension and number of compute nodes.
        (a)~Increasing node count consistently reduces simulation time.
        (b)~Additional nodes efficiently distribute computational workload.
        (c)~Communication costs are non-negligible and a bottleneck in our current implementation.
    }
    \label{fig:all_three_plots}
\end{figure*}

We conducted the benchmark simulations on the Perlmutter supercomputer. Each compute node on this system contains four NVIDIA A100 GPUs. Each node in our distributed scheme functions as a single rank.

The overall performance of the simulator is illustrated in Figure 6. This figure shows the total simulation time for the brickwork circuit as a function of the qubit count, with a fixed maximum bond dimension of 250 and a depth of 80 layers. The runtime scales linearly with the number of qubits, an expected result for an MPS based simulation. The results also demonstrate that increasing the number of nodes from one to four effectively reduces the total simulation time. The diminishing returns with more nodes, however, suggest the presence of a significant overhead that we investigate further.

We present a detailed breakdown of the execution time for a 64 qubit circuit in Figure 7. This analysis separates the total runtime into its two main components. These components are computation, including tensor contraction and SVD, and communication, including synchronization and data movement. The data in Figure 7(b) confirms that the computational workload is distributed effectively. The computation time scales almost perfectly inversely with the number of nodes. In contrast, Figure 7(c) reveals that synchronization and data movement costs constitute a non negligible bottleneck. While computation time decreases, this communication overhead begins to dominate the total runtime. This effect limits the overall scalability as more nodes are added.

This trade off between parallel computation and communication overhead is the primary motivation for our hybrid parallelization architecture. For circuits with high gate density per layer but moderate bond dimensions, our task based approach efficiently distributes the work. The initial synchronization overhead is tolerable and allows for rapid execution. The advantage of this architecture is its flexibility. For simulations where the bond dimension grows exceedingly large, individual tensor contractions become the bottleneck. The framework can then be extended to allow multiple GPUs per subgroup to collaborate on a single contraction using TAMM's block level data distribution. This approach mitigates the challenges of distributed tensor contraction schemes.

\section{Conclusion}

We have demonstrated the integration of a tensor network simulation backend into the NWQ-Sim software package, leveraging the TAMM library to facilitate scalable and efficient tensor algebra operations across diverse GPU architectures. Although initial development and evaluation were conducted using NVIDIA GPUs on the Perlmutter supercomputer, the modular and portable design of the software provides a clear path for enabling compilation and execution on other leading HPC platforms, such as Frontier (AMD) and Aurora (Intel). To utilize GPU-accelerated SVD on these platforms, for example, \verb|rocsolver_cgesvd| would be used for AMD GPUs and \verb|gesvd| for Intel GPUs within the TAMM backend.

Building upon the foundational capabilities established here, several avenues for further optimization and enhancement have been identified. Advanced tensor network contraction algorithms, particularly those designed to optimize contraction sequences to balance memory efficiency, computational error, and runtime performance, represent a promising direction for continued improvement \cite{b26}. Techniques such as hyperoptimization of approximate tensor network contractions provide compelling opportunities to enhance simulator scalability and computational efficiency~\cite{b30}.

The system presented in this paper thus establishes a robust baseline for a highly configurable, hardware-agnostic, and performant tensor network simulation environment. Looking forward, we plan to incorporate additional tensor network topologies, including Projected Entangled Pair States (PEPS) and Tree Tensor Networks (TTN). These different topologies are better suited to simulating the entanglement structures present in various physical quantum computing architectures (e.g., 2D grids of superconducting qubits) and for specific types of quantum simulation algorithms. 

The current implementation of the parallelized TAMM backend allocates a single GPU per subgroup. In the future a potential avenue to explore is dynamically changing the group size as gate contraction becomes too large for a single GPU. This would require an update to tensor decomposition, which is currently done locally per GPU with the NVIDIA \verb|cusolverDnZgesvdj| function. Future iterations would employ parallelized SVD schemes such as MAGMA's \verb|zgesvd|. A dynamic Tensor Network backend would be able to be efficient in wide and highly parallel circuits as well as circuits with high bond dimension requiring parallelized contraction.

Finally, the current implementation of TN-Sim focuses exclusively on pure-state MPS simulation, which inherently limits its effectiveness in modeling quantum noise processes. To overcome this limitation, future work will extend the framework by integrating matrix product operators (MPOs), enabling efficient representation of higher-dimensional density matrices. This enhancement will facilitate detailed simulation and analysis of quantum error channels, thus supporting comprehensive studies of error correction and mitigation strategies essential for practical quantum computation.

\section*{Acknowledgment}

This material is based upon work supported by the U.S. Department of Energy, Office of Science, National Quantum Information Science Research Centers, Quantum Science Center. This material is based upon work supported by the National Science Foundation under Award No. 2021540 (Accelerating Quantum‑Enabled Technologies). This research used resources of the National Energy Research Scientific Computing Center (NERSC), a Department of Energy User Facility using NERSC award DOE-ERCAP0032559. This research used resources of the Oak Ridge Leadership Computing Facility, which is a DOE Office of Science User Facility supported under Contract DE-AC05-00OR22725. The Pacific Northwest National Laboratory is operated by Battelle for the U.S. Department of Energy under Contract DE-AC05-76RL01830.

We thank Ajay Panyala for valuable discussions that contributed to the development of this work.

\end{document}